\documentclass[aps,twocolumn]{revtex4-1}
\usepackage{graphicx}
\usepackage[justification=justified,width=\linewidth]{caption}
\usepackage{subcaption}
\usepackage{amsmath}
\usepackage{amsfonts}
\usepackage{amsthm}
\usepackage{amssymb}
\usepackage{amsbsy}
\usepackage{wasysym}
\usepackage{bm}
\usepackage{mathrsfs}
\usepackage{color}
\usepackage{times}
\usepackage[resetlabels]{multibib}

\begin{document}

\title{Active alignment-driven coarsening in confined near-critical fluids} 
\author {Parameshwaran A and Bhaskar Sen Gupta}
\email{bhaskar.sengupta@vit.ac.in}
\affiliation{Department of Physics, School of Advanced Sciences, Vellore Institute of Technology, Vellore, Tamil Nadu - 632014, India}

\date{\today}
            
\begin{abstract}
We investigate vapor-liquid phase separation of an active near critical Lennard-Jones fluid confined within a cylindrical pore using molecular dynamics simulations. Activity is introduced via Vicsek-type alignment interactions, enabling a systematic study of how	self-propulsion modifies domain morphology and coarsening kinetics under
quasi-one-dimensional confinement. In the passive limit, the system undergoes early-time spinodal decomposition (diffusive growth characterized by the Lifshitz-Slyozov exponent $\alpha = 1/3$), followed by the formation of periodically modulated, plug-like liquid domains along the pore axis. At late times, coarsening becomes kinetically arrested, and the system remains trapped in a metastable striped state. Introducing activity destabilizes this arrested morphology by enhancing collective domain transport, leading to frequent domain mergers and complete phase separation at sufficiently high activity. The late-stage coarsening then exhibits a crossover to faster, ballistic growth with an effective exponent $\alpha = 2/3$, consistent with a cluster-coalescence mechanism. Analysis of two-point correlation functions and structure factors confirms dynamic scaling across all activity regimes. Our results demonstrate that alignment-induced activity can overcome confinement-driven kinetic arrest, providing new insight into phase separation in confined active fluids. The relevant growth laws are analyzed and interpreted using appropriate theoretical frameworks.
\end{abstract}
\maketitle
\section{Introduction}
Phase separation in fluids is a paradigmatic nonequilibrium process that has been extensively studied in the context of vapor-liquid and liquid-liquid transitions (see~\cite{suman} and the references therein). Following a quench from a homogeneous high-temperature state into the two-phase region, the system undergoes domain formation and coarsening, characterized by universal growth laws that depend on conservation laws, transport mechanisms, and dimensionality \cite{puri,Onuki,Binder,Jones}. While such processes are well understood in bulk equilibrium systems, their behavior can be profoundly altered under confinement, where geometric constraints introduce new length scales and may strongly suppress hydrodynamic transport~\cite{Preethi,Bhattacharyya-1}.

Confinement-induced modifications of phase separation are particularly relevant in porous and quasi-one-dimensional geometries, such as nanopores, microfluidic channels, and porous membranes~\cite{Liu-1,Liu-2,Tananka,Parameshwaran-1}. In these systems,  phase separation often proceeds through the formation of axially modulated, plug-like domains rather than bulk bicontinuous morphologies. At late times, coarsening may become kinetically arrested, resulting in long-lived metastable striped states due to reduced connectivity and weakened interfacial interactions along the confined direction~\cite{daniya}. Understanding the interplay between confinement, morphology, and coarsening kinetics is therefore essential for applications ranging from fluid transport in porous media to controlled pattern formation in micro- and nanofluidic devices~\cite{Wang}.

In parallel, the last decade has witnessed rapid progress in the study of active matter systems, composed of self-driven units that continuously consume energy at the microscopic scale \cite{Ramaswamy,Marchetti,Rao,Cates2015}. Active fluids exhibit rich collective phenomena, including clustering, motility-induced phase separation, and anomalous transport, which have no equilibrium counterpart. Experimental realizations span a wide range of systems, including bacterial suspensions, synthetic Janus colloids, cytoskeletal filaments, and active granular media \cite{Schaller2010, Bechinger2016}. Of particular interest is how activity modifies classical phase separation kinetics, often leading to enhanced transport, altered scaling behavior, and departures from equilibrium universality classes.

Many biological systems, e.g. motile bacteria, active colloids, membraneless organelles exhibit features reminiscent of active vapor-liquid phase separation, where motor proteins, cytoskeletal forces, and biochemical activity sustain growth, mobility, and turnover of condensates~\cite{breg,zottl}. Despite growing interest, the kinetics of vapor-liquid phase separation of near critical active fluids (density close to the critical value) under confinement remains relatively unexplored.  A systematic understanding of how activity competes with confinement-induced kinetic arrest in near critical fluids, particularly in quasi-one-dimensional pores relevant to simple porous media, is still lacking. In particular, it is unclear whether activity can restore hydrodynamic-like growth under confinement and whether the resulting coarsening obeys universal scaling laws.

In this work, we address these questions by studying vapor-liquid phase separation of an active Lennard-Jones fluid confined inside a cylindrical pore using molecular dynamics simulations. Activity is introduced through Vicsek-type alignment interactions, which generate coherent collective motion without directly modifying the interparticle potential~\cite{vicsek1995,vicsek2000}. This allows us to isolate the effect of alignment-induced activity on morphology and kinetics under strong confinement. By systematically varying the activity strength, we investigate the transition from kinetically arrested, diffusion-dominated coarsening in the passive limit to faster, activity-driven growth at higher self-propulsion.

We characterize the evolving morphologies using real-space snapshots, two-point correlation functions, and structure factors, and extract the characteristic domain size to quantify coarsening dynamics. Our results reveal a clear activity-induced crossover in the growth law, from Lifshitz-Slyozov scaling to ballistic cluster coalescence, accompanied by systematic changes in domain morphology and scaling functions. These findings demonstrate how activity can overcome confinement-induced kinetic arrest and provide new insights into phase separation in confined active fluids, with direct relevance to transport and pattern formation in porous media.

\section{Model and Methods}
We consider a single-component vapor-liquid system confined within a cylindrical nanopore. 
The passive interaction between a pair of particles $i$ and $j$, separated by a distance 
$r_{ij}=|\vec{r}_i-\vec{r}_j|$, is described by the standard Lennard-Jones (LJ) potential,
\begin{equation}
	U(r_{ij})=4\epsilon\left[\left(\frac{\sigma}{r_{ij}}\right)^{12}
	- \left(\frac{\sigma}{r_{ij}}\right)^6\right].
	\label{eq1}
\end{equation}
Here, $\sigma$ and $\epsilon$ denote the particle diameter and interaction strength, respectively. Throughout this work, we employ reduced LJ units by setting $\sigma=\epsilon=m=k_B=1$. The length, temperature and time are scaled in units of $\sigma$, $\epsilon/k_B$ and $(m\sigma^2/\epsilon)^{1/2}$ respectively. To enhance computational efficiency, the potential is truncated at a cutoff distance  $r_c=2.5\sigma$. Since a simple truncation introduces discontinuities in both the potential and force, we use a shifted-force LJ potential defined as
\begin{equation}
	V(r_{ij}) = \begin{cases}
		\text{if } (r_{ij} < r_c); \vspace{0.05cm}\\
		U(r_{ij}) - U(r_c) - (r_{ij} - r_c) \left(\frac{dU}{dr_{ij}}\right)_{r_{ij}=r_c}, \vspace{0.2cm}\\
		
		\text{if } (r_{ij} \geq r_c) ; \hspace{1cm}0 .
	\end{cases}
\end{equation}
This construction ensures continuity of both the potential and the force at the cutoff. The passive LJ system described above exhibits a bulk vapor-liquid critical temperature 
$T_c=0.94\,\epsilon/k_B$ and a critical number density 
$\rho_c=N/V=0.32$~\cite{roy2012nucleation,daniya-1,paramesh-2}. Here $N$ and $V$ represent the number of particles and the volume of the simulation box respectively. 

\subsection{Confinement Geometry}
\begin{figure}[h]
	\centering
	\includegraphics[width=0.4\textwidth]{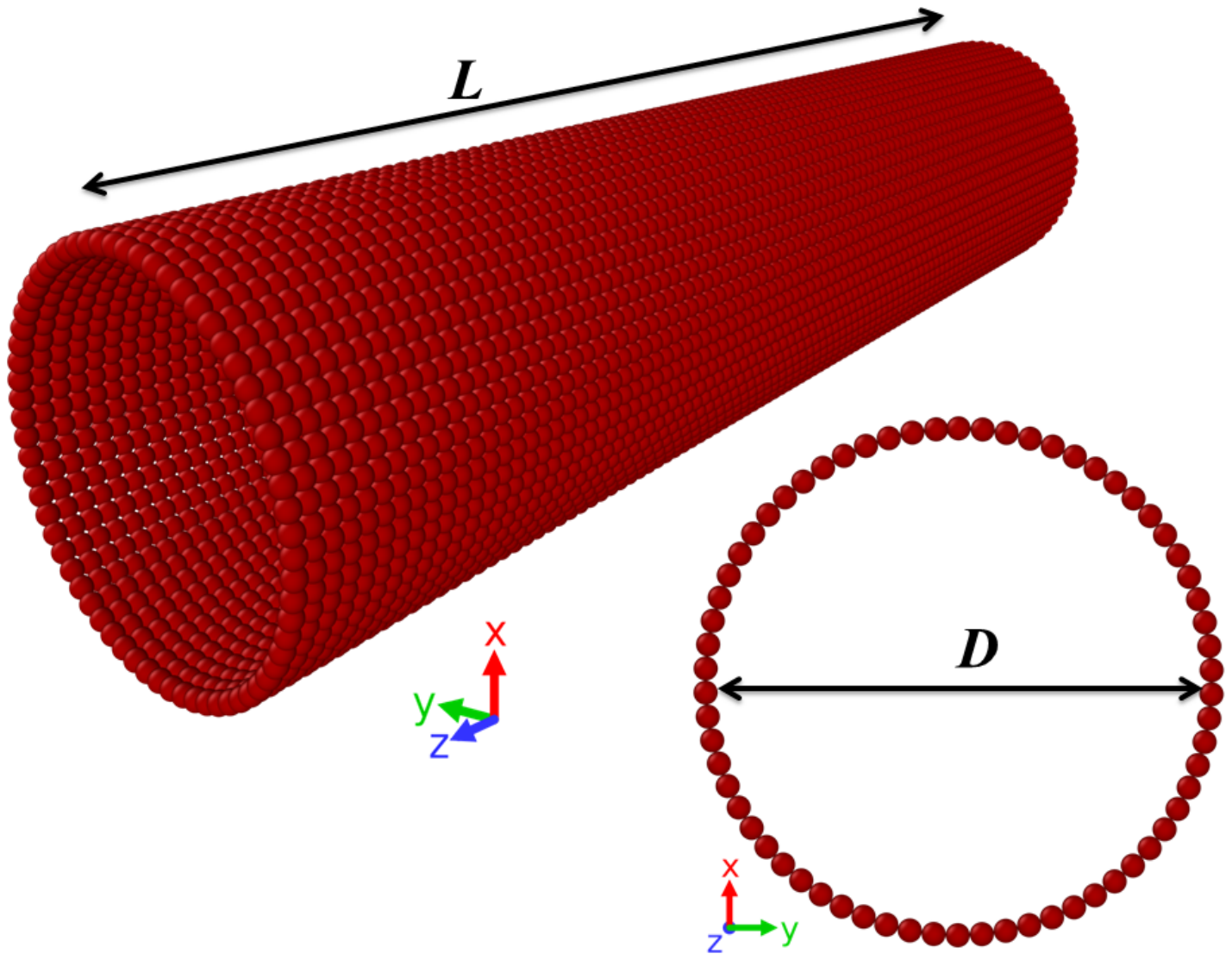}
	\caption{Schematic of the cylindrical nanopore of length $L$ and diameter $D$, with periodic boundary conditions along $z$. Walls are composed of immobile particles at density $\rho_w=1$.}
	\label{fig:cylinder}
\end{figure}

The cylindrical nanopore is modeled as a tube of length $L$ and diameter $D$, with  $L\gg D$, as illustrated schematically in Fig.~\ref{fig:cylinder}. Periodic boundary conditions are imposed along the axial ($z$) direction, effectively mimicking an infinitely long pore.
The pore walls are constructed from immobile LJ particles of number density  $\rho_w=1$ and diameter $\sigma$, identical to that of the fluid particles. The LJ interaction between wall and fluid particles is truncated at $r_c=2^{\frac{1}{6}}\sigma$, ensuring an effectively neutral confinement.
\subsection{Vicsek-Type Activity}
Self-propelled activity is introduced through a Vicsek-like alignment force~\cite{vicsek1995,vicsek2000}.
Each particle experiences an active force
\begin{equation}
	\vec{f}_i = f_A \hat{D},
\end{equation}
where $f_A$ denotes the activity strength and $\hat{D}$ is the local mean velocity 
direction of neighboring particles within a sphere of radius $r_c=2.5\sigma$,
\begin{equation}
	\hat{D} = \frac{\sum_j \vec{v}_j}{\left|\sum_j \vec{v}_j\right|}.
\end{equation}
Here, the sum runs over all neighboring particles $j$ within the interaction range.

A naive implementation of this force would modify both the direction and magnitude of 
particle velocities, leading to uncontrolled heating and a shift in the effective 
temperature. We therefore modify the velocity update rule so that  the active force affects only the direction of motion while preserving the magnitude of the passive velocity. Specifically, the updated velocity is written as~\cite{das2017}
\begin{equation}
	\vec{v}_i(t+\Delta t) = 
	\left|\vec{v}_i^{\mathrm{pas}}(t+\Delta t)\right| \hat{n},
\end{equation}
where
\begin{equation}
	\hat{n} =
	\frac{\vec{v}_i^{\mathrm{pas}}(t+\Delta t) + 
		\dfrac{\vec{f}_i}{m_i}\Delta t}
	{\left|\vec{v}_i^{\mathrm{pas}}(t+\Delta t) + 
		\dfrac{\vec{f}_i}{m_i}\Delta t\right|}.
\end{equation}
Here, $\vec{v}_i^{\mathrm{pas}}(t+\Delta t)$ denotes the velocity obtained from passive 
interactions alone. This implementation closely resembles the original Vicsek dynamics, 
while allowing independent control over the system temperature.

\subsection{Simulation Details}

Molecular dynamics simulations are performed in the canonical ensemble using an in-house  OPENMP-parallelized Fortran code. Temperature is maintained using a Langevin thermostat,  with equations of motion given by~\cite{allen1987,frenkel2002}
\begin{equation}
	m_i \ddot{\vec{r}}_i =
	- \vec{\nabla} V_i
	- \gamma m_i \dot{\vec{r}}_i
	+ \sqrt{2\gamma k_B T m_i}\,\vec{\eta}_i(t)
	+ \vec{f}_i,
\end{equation}
where $\gamma$ is the translational damping coefficient and $\vec{\eta}_i(t)$ is a Gaussian  white noise satisfying
\begin{equation}
	\langle \eta_{i\mu}(t)\eta_{j\nu}(t') \rangle =
	\delta_{ij}\delta_{\mu\nu}\delta(t-t').
\end{equation}

Hydrodynamics-preserving thermostats are deliberately avoided in order to isolate the effects of Vicsek-type activity. Consequently, in the passive limit ($f_A=0$), the dynamics  corresponds to Model~B~\cite{puri}. All simulations are carried out using the velocity-Verlet integration scheme~\cite{Verlet} with time step $\Delta t = 0.01\tau$, where $\tau=(m\sigma^2/\epsilon)^{1/2}=1$, ensuring numerical stability.

Unless stated otherwise, results are reported for $L=500\sigma$, $D=20\sigma$,  temperature $T=0.6\,\epsilon/k_B$, and number density $\rho=0.3$. Initial configurations  are equilibrated at a high temperature $T=6.0\,\epsilon/k_B$ in the passive limit  ($f_A=0$), followed by an instantaneous quench to $T=0.6\,\epsilon/k_B$. The subsequent  phase separation dynamics is studied for activity strengths in the range  $f_A=0$-$0.8$. All reported results are averaged over 30 independent initial  configurations.


\section{Results}

\begin{figure}[h]
	\centering
	\includegraphics[width=0.4\textwidth]{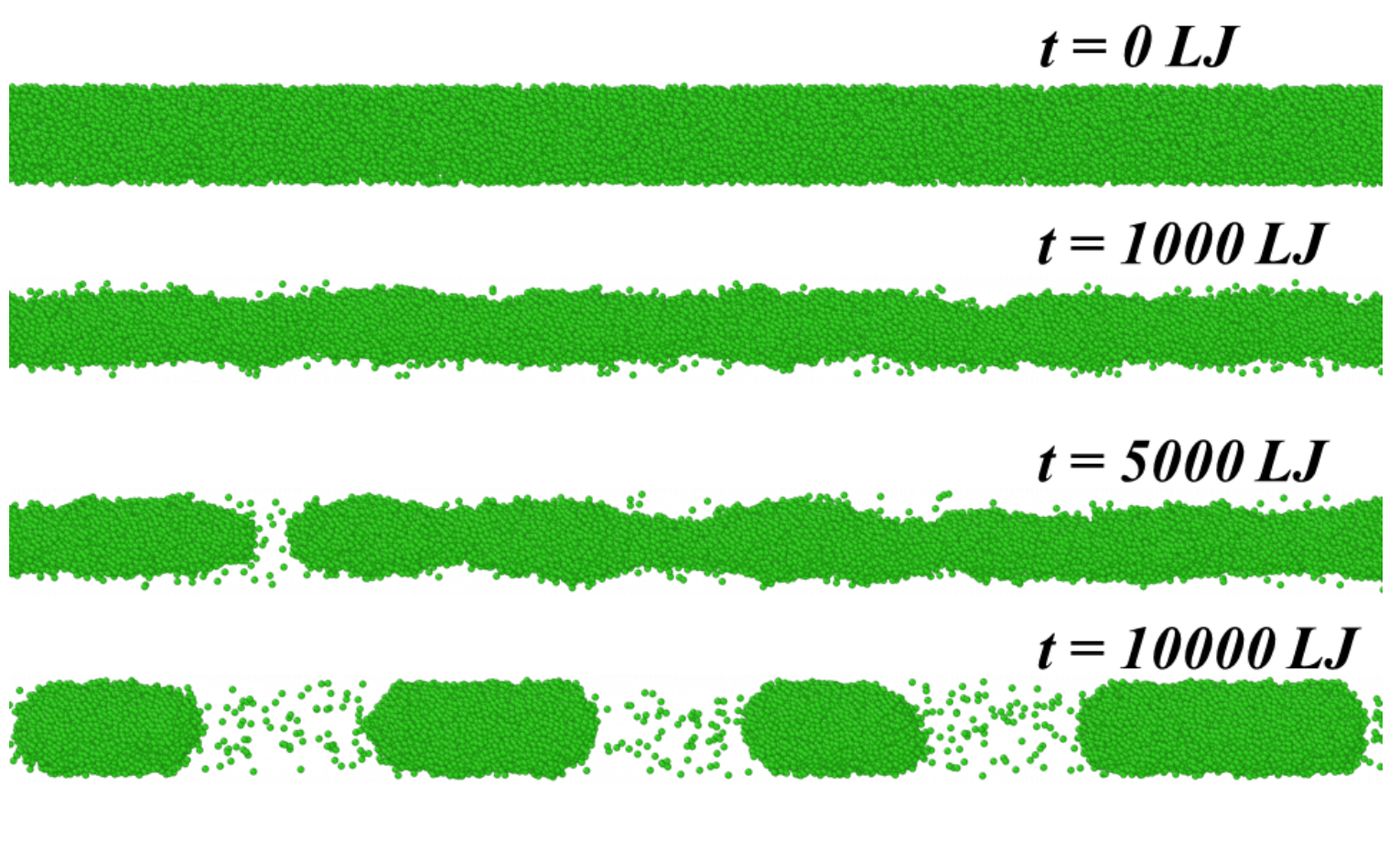}
	\caption{Time evolution of the passive system ($f_A$=0) following a quench to $T=0.6<T_c$ at density $\rho=0.3$. Green dots represent particles. Cylindrical confinement leads to spinodal decomposition and the formation of axially modulated plug-like liquid domains.}
	\label{fig:Ev_f0}
\end{figure}

Figure~\ref{fig:Ev_f0} shows the time evolution of a cylindrically confined vapor-liquid system in the passive limit ($f_A=0$) at an overall number density $\rho=0.3$, close to the criticality. In unconfined bulk vapor-liquid systems, a sudden quench from a high-temperature homogeneous phase typically leads to spinodal decomposition, characterized by the formation of interconnected liquid and vapor domains prior to complete phase separation~\cite{suman}. A similar early-stage interconnected morphology is observed here under cylindrical confinement.

As the system evolves, the confinement geometry strongly influences the domain morphology. In particular, the initially interconnected structure reorganizes into periodically spaced, plug-like liquid domains aligned along the axis of the cylindrical pore, as shown in Fig.~\ref{fig:Ev_f0}. Such morphologies arise from the tendency of the system to minimize interfacial free energy under quasi-one-dimensional confinement conditions~\cite{Liu-1,Liu-2}. In the passive case, the widening of these periodic plug-like domains ceases at late times, and the system becomes trapped in a striped metastable state. This arrest can be attributed to the absence of effective bridging events between neighboring domains along the pore axis, as well as to weak interactions between distant interfaces once their separation exceeds a characteristic length scale~\cite{daniya}.

\begin{figure}[h]
	\centering
	\includegraphics[width=0.4\textwidth]{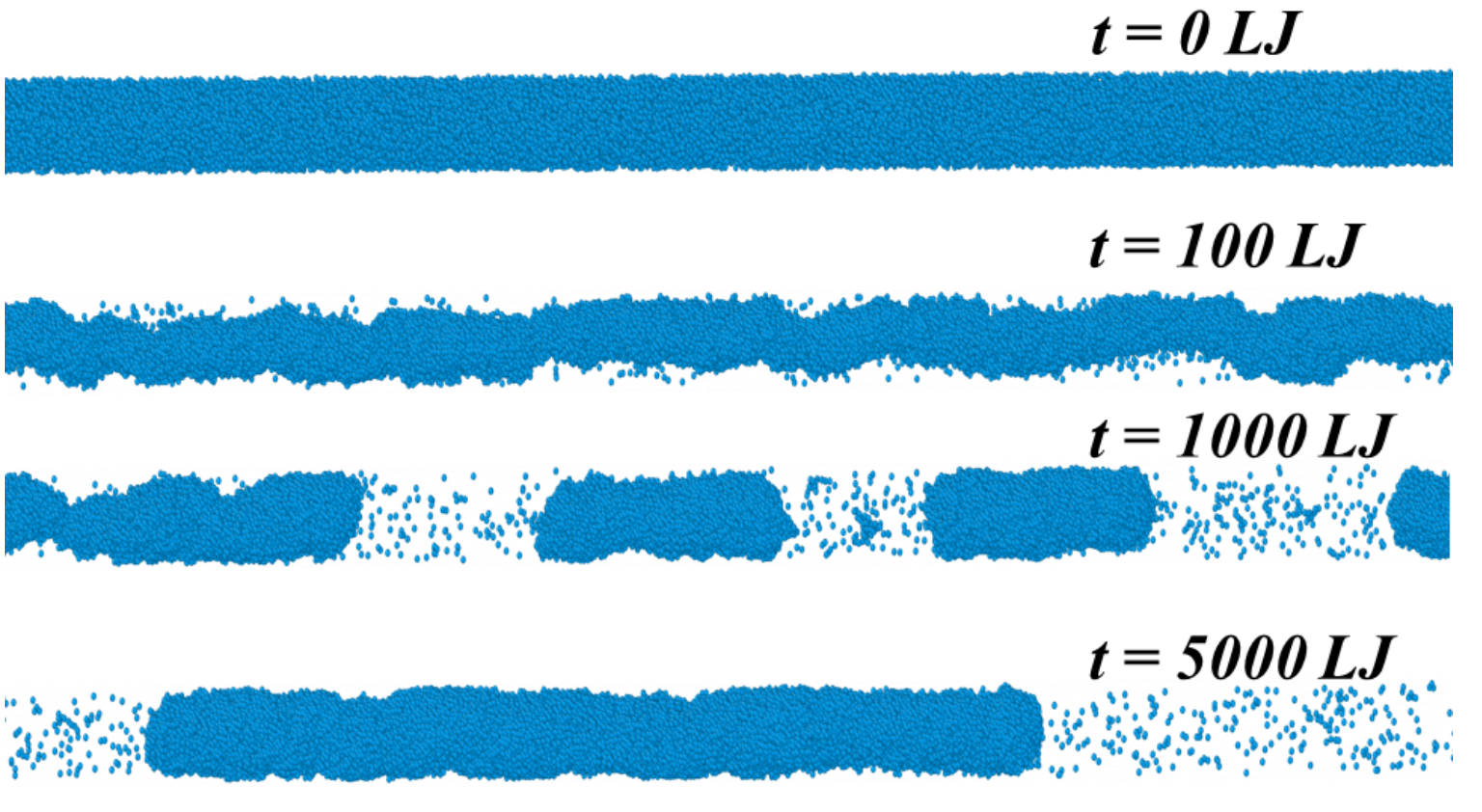}
	\caption{Time evolution of the active system ($f_A=0.8$) after a quench to $T=0.6$, at $\rho=0.3$. Blue dots represent particles. Activity induces rapid coarsening and promotes axial domain motion and mergers.}
	\label{fig:f08}
\end{figure}

The introduction of Vicsek-type activity qualitatively alters the coarsening dynamics.
We systematically investigate the effect of alignment-driven self-propulsion on domain
growth over a wide range of activity strengths, $0 \le f_A \le 0.8$. Figure~\ref{fig:f08}
shows the temporal evolution of the system for a strongly active case ($f_A=0.8$). In
contrast to the passive limit, coarsening in the active system proceeds significantly
faster.

\begin{figure}[h]
	\centering
	\includegraphics[width=0.4\textwidth]{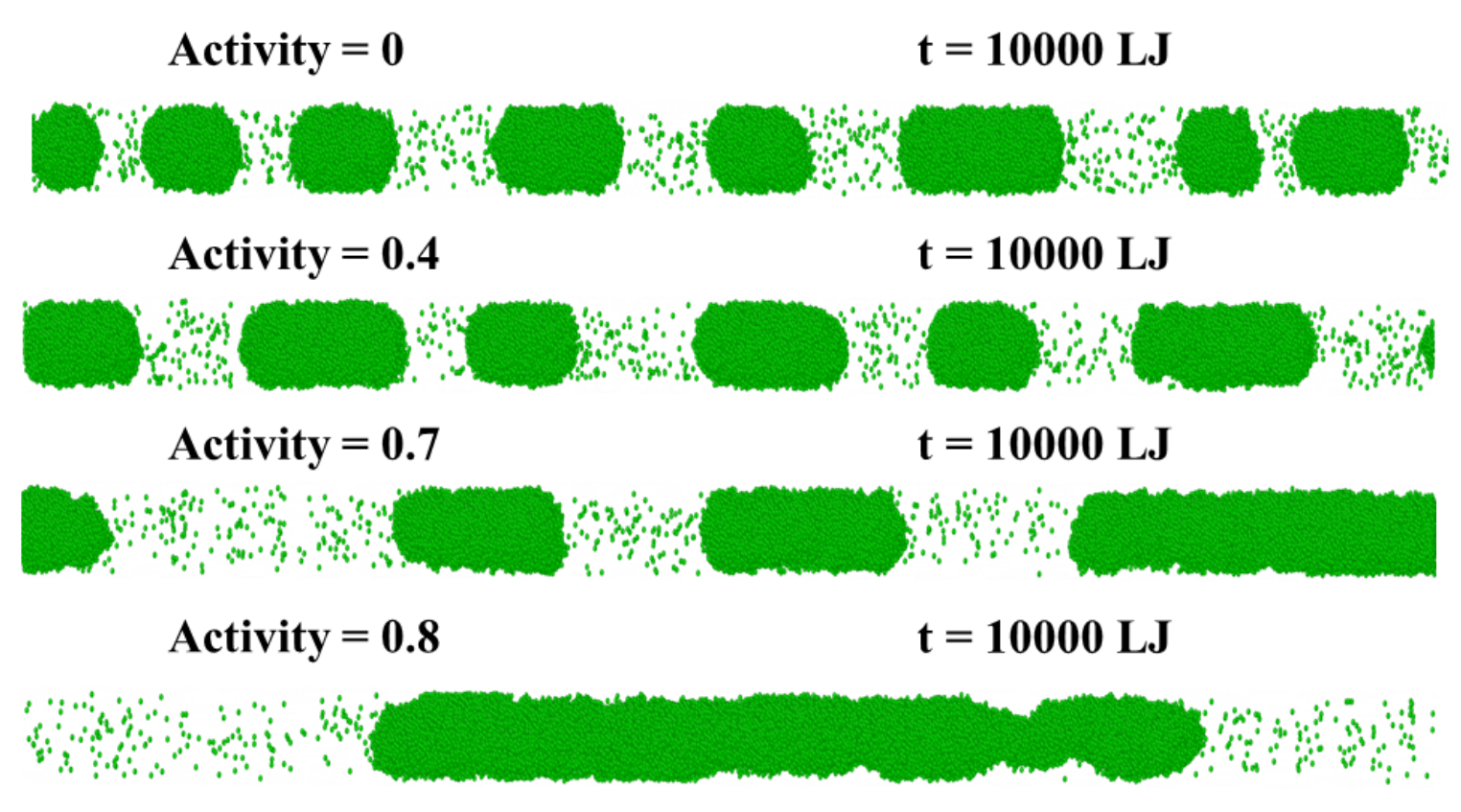}
	\caption{Late-time configurations ($t = 10^4$) for different activity strengths ($0 \le f_A \le 0.8$) at $T = 0.6$ and $\rho = 0.3$. Increasing activity enhances axial mobility and promotes domain coalescence.}
	\label{fig:fig4}
\end{figure}

The accelerated growth can be attributed to the coherent motion induced by Vicsek-like
alignment interactions, which generates an effective advection-like transport of particles and clusters~\cite{paul2017ballistic}. As the activity strength increases, the late-stage periodic plug-like domains acquire substantial mobility along the pore axis and undergo repeated collisions and mergers. This is clearly illustrated in Fig.~\ref{fig:fig4}, where increasing activity leads to progressively larger domains and ultimately to the breakdown of the metastable striped state. For sufficiently strong activity ($f_A=0.8$), complete phase separation becomes possible within the cylindrical confinement.

To quantify the enhanced mobility of domains, we compute the mean-squared displacement (MSD) of the center of mass (CM) of late-stage plug-like clusters~\cite{hansen},
\begin{equation}
	\mathrm{MSD}_{\mathrm{CM}}(t) =
	\left\langle
	\left[ \vec{R}_{\mathrm{CM}}(t) - \vec{R}_{\mathrm{CM}}(0) \right]^2
	\right\rangle,
\end{equation}
where the cluster center of mass is defined as
\begin{equation}
	\vec{R}_{\mathrm{CM}} =
	\frac{1}{M_c}\sum_{i=1}^{N_c} m_i \vec{r}_i.
\end{equation}
Here, $M_c$ and $N_c$ denote the mass and number of particles in the cluster,
respectively. The time origin is chosen at the onset of the MSD measurement and does
not correspond to the initial simulation time.

\begin{figure}[h]
	\centering
	\includegraphics[width=0.4\textwidth]{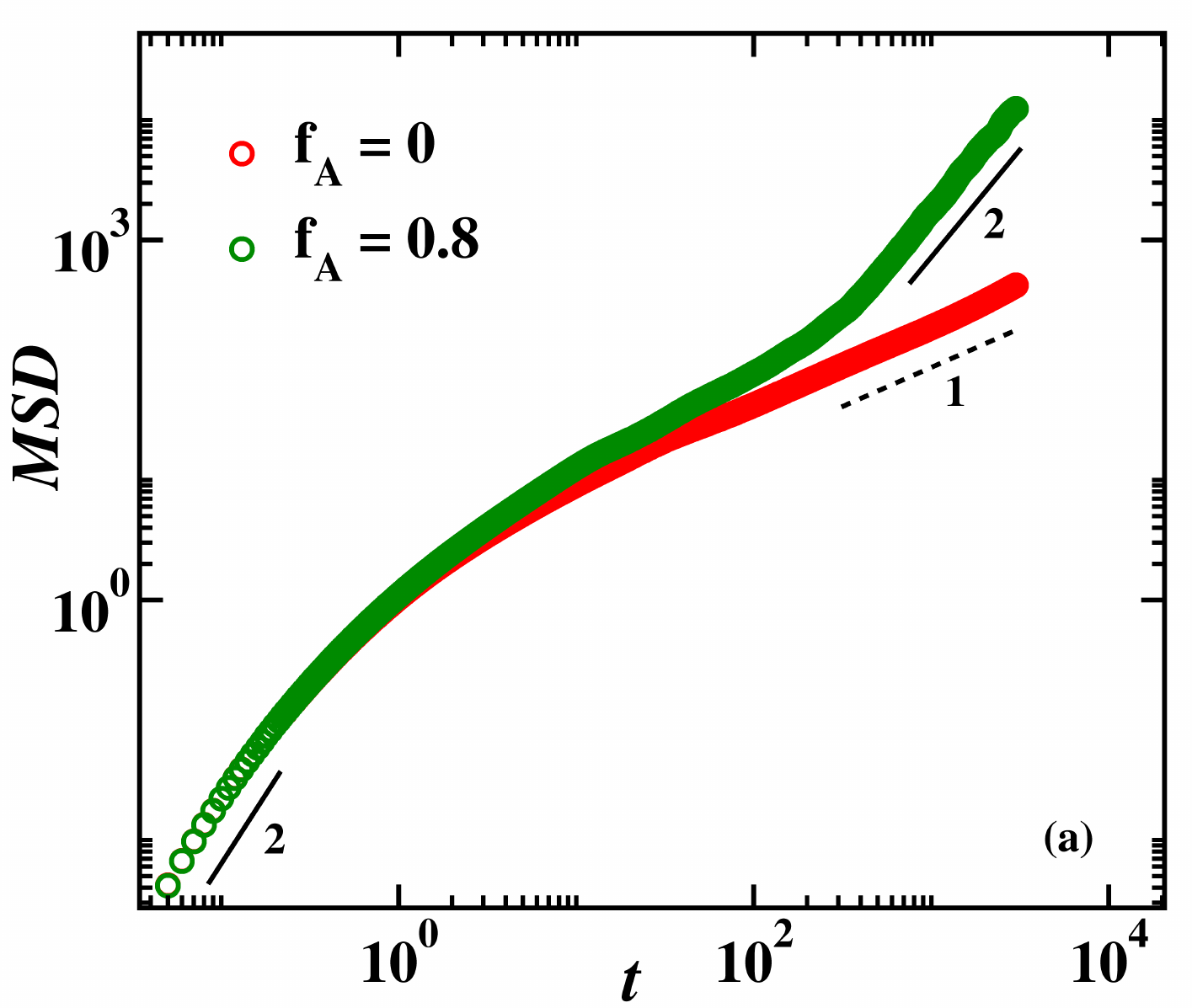}
	\includegraphics[width=0.4\textwidth]{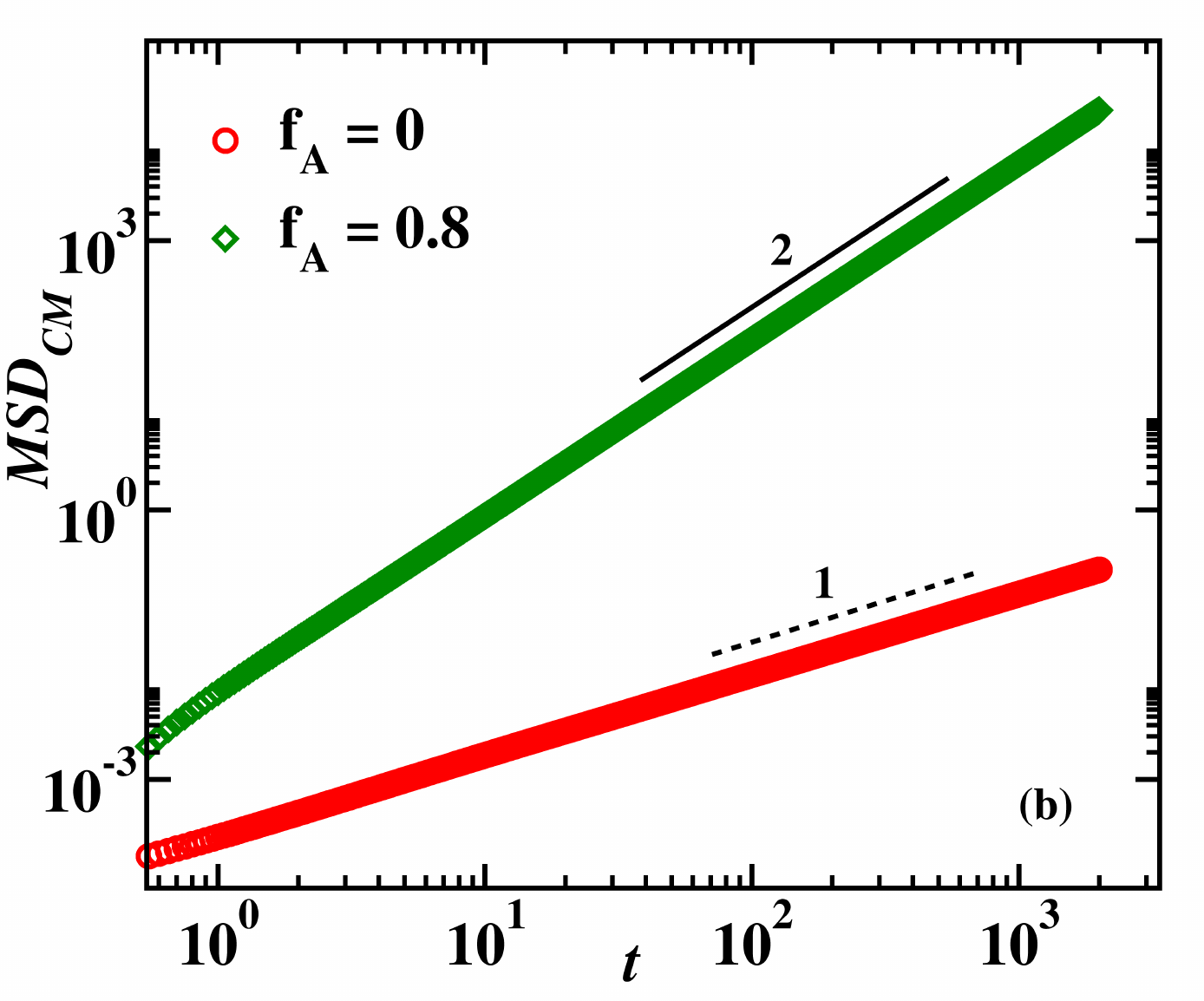}
	\caption{ The mean-squared displacement of (a) individual particles for $f_A=0$ and
		$f_A=0.8$, (b) the center of mass of clusters for the same activity values. Dotted lines indicate diffusive scaling ($t^1$), while solid lines indicate ballistic scaling ($t^2$).}
	\label{fig:C_Msd}
\end{figure}

Figure~\ref{fig:C_Msd}(a) compares the average individual particle MSD for the active and passive cases. At short times, both systems display ballistic motion, which then transitions to a conventional diffusive regime. At intermediate times, however, the active system deviates toward a superdiffusive behavior. At longer times, the Vicsek activity leads to a sustained ballistic regime.

Figure~\ref{fig:C_Msd}(b) presents the average MSD of the clusters. The results reveal a substantial difference between the two systems: the MSD in the passive case is about two orders of magnitude lower than in the active case. Consequently, the passive clusters exhibit negligible motion and remain effectively stationary.  In contrast, for the active case ($f_A=0.8$), the MSD exhibits a clear ballistic regime,
$\mathrm{MSD}_{\mathrm{CM}} \sim t^2$, indicating persistent directed motion of clusters. This behavior is consistent with a ballistic cluster coalescence mechanism~\cite{paul2017ballistic,carnevale1990,trizac2003}, which enables efficient domain merging along the cylindrical axis.

To characterize the evolving domain morphology quantitatively, we compute the two-point equal-time correlation function along the axial direction~\cite{Parameshwaran-1,daniya,daniya-2},
\begin{equation}
	C(z,t) =
	\langle \psi(0,t)\psi(z,t) \rangle
	- \langle \psi(0,t)\rangle \langle \psi(z,t)\rangle,
\end{equation}
where $\psi(z,t)$ is the local order parameter. The angular brackets represent the ensemble averaging. Owing to the strong geometrical constraints imposed by cylindrical confinement, we define $\psi(z,t)$ by dividing the system into axial slices of width $1.0\sigma$. A slice is assigned $\psi=+1$ if its local number density exceeds $\rho_c$, and $\psi=-1$ otherwise~\cite{Parameshwaran-1,daniya}.

For statistically self-similar growth, the correlation function is expected to obey dynamic scaling~\cite{bray2002},
\begin{equation}
	C(z,t) \equiv \tilde{C}\!\left( \frac{z}{\ell(t)} \right),
\end{equation}
where $\ell(t)$ is a characteristic length scale, representing the average linear size of domains along the axial direction, extracted from the first zero crossing of $C(z,t)$.

\begin{figure}[h]
	\centering
	\includegraphics[width=0.147\textwidth]{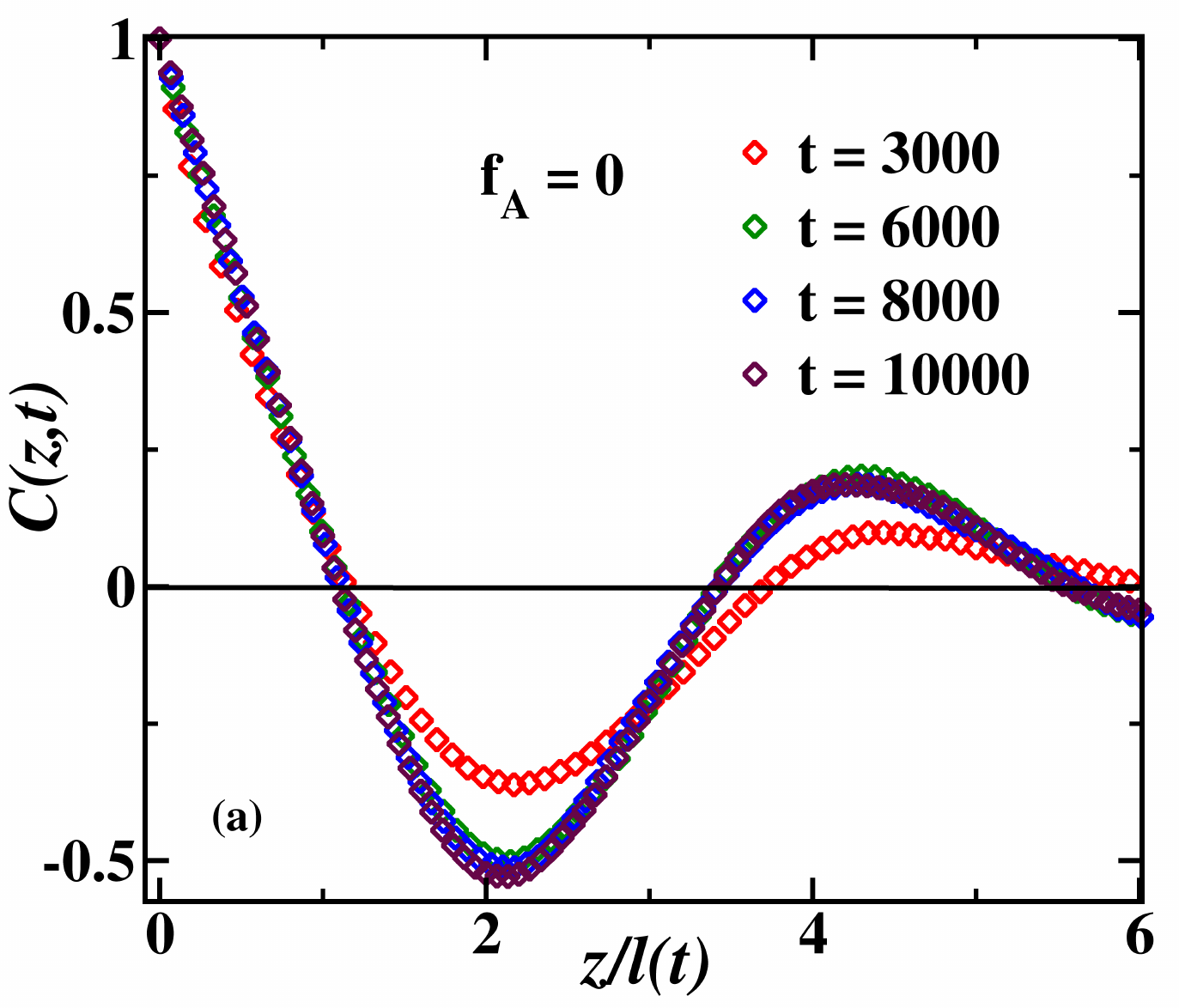}
	\includegraphics[width=0.15\textwidth]{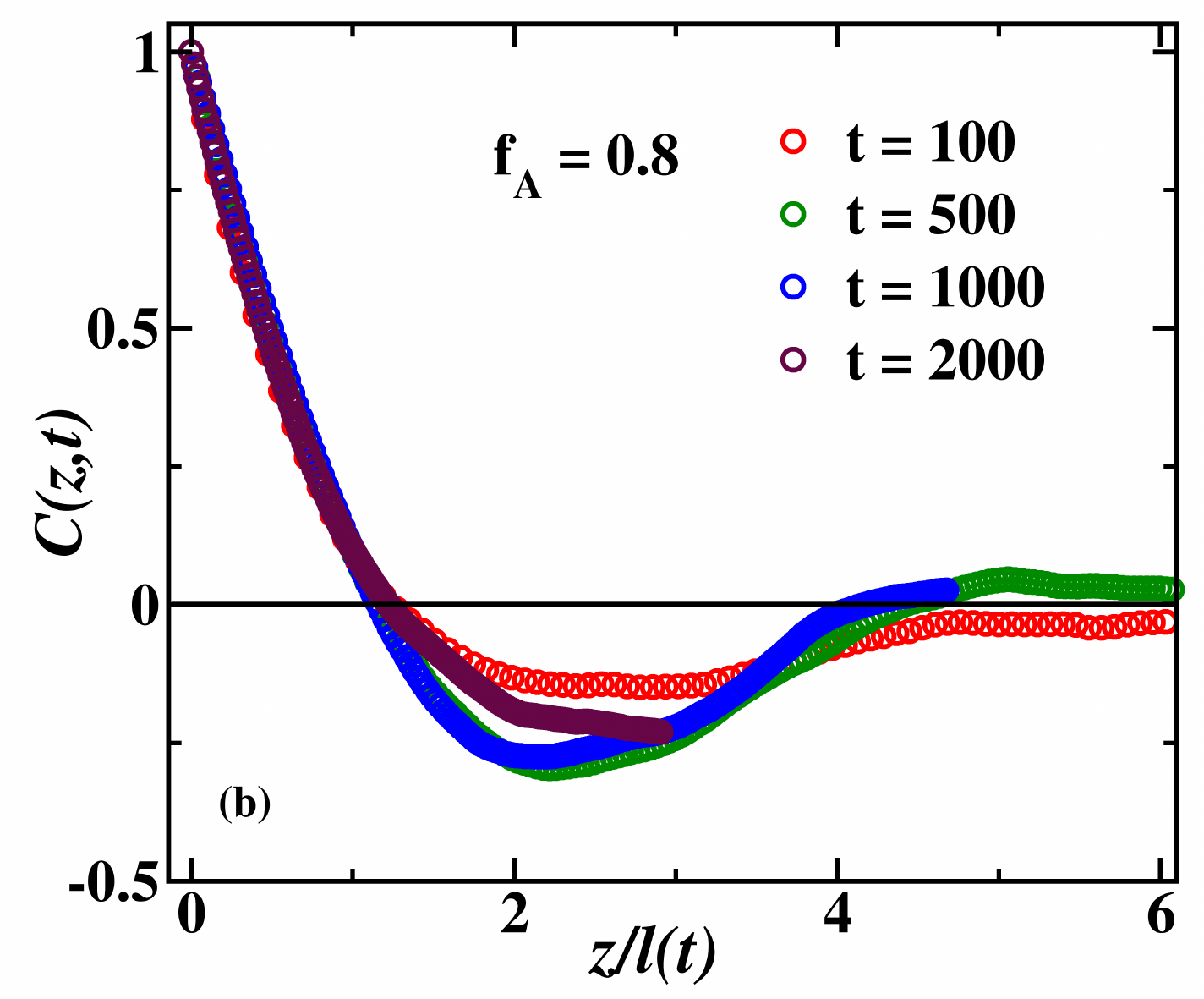}
    \includegraphics[width=0.137\textwidth]{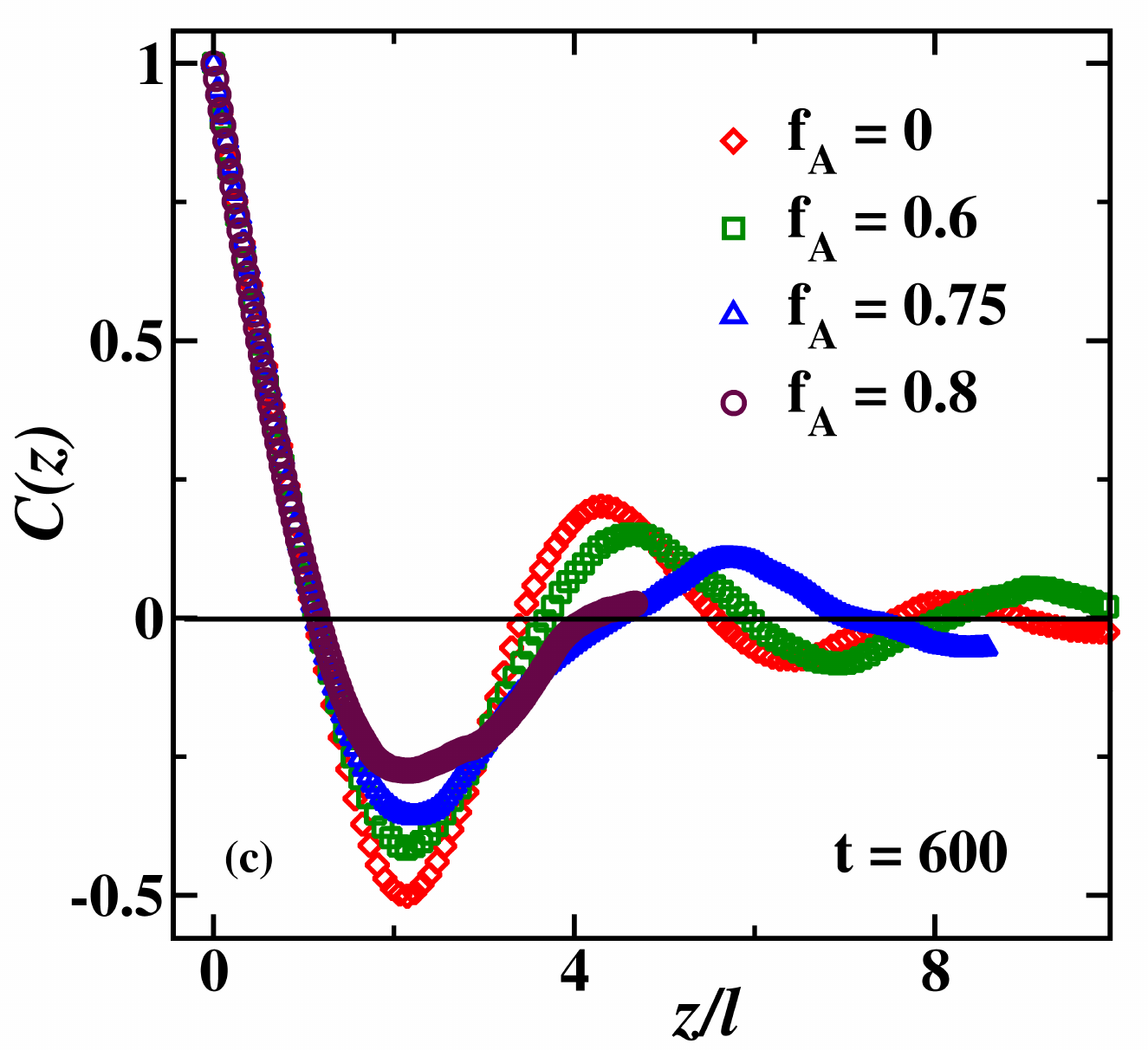}
	\caption{Scaling collapse of the correlation function for (a) passive case ($f_A=0.0$), (b) active case ($f_A=0.8$) and (c) across different activity strengths.}
	\label{fig:C_cor1}
\end{figure}

Figure~\ref{fig:C_cor1}(a) and ~\ref{fig:C_cor1}(b) demonstrate excellent scaling collapse of $C(z,t)$ against $z/\ell(t)$  over multiple times for both the passive limit  and the active case (only $f_A=0.8$ is shown), confirming self-similar growth within each case. Next we compare the active and passive cases. Figure~\ref{fig:C_cor1}(c) shows the scaled correlation functions for different activity strengths do collapse onto a universal curve, indicating that superuniversality~\cite{su1,su2} remains vindicated for active fluids under confinement.

Further insight into interfacial morphology is obtained from the structure factor, defined as the Fourier transform of $C(z,t)$~\cite{bray2002,rounak-1,rounak-2},
\begin{equation}
	S(k_z,t) = \langle |\psi(k_z,t)|^2 \rangle
	= \int e^{ikz} C(z,t)\,dz.
\end{equation}
For self-similar growth, $S(k_z,t)$ satisfies the scaling form
\begin{equation}
	S(k_z,t) \equiv \ell(t)^d \tilde{S}(k_z\ell(t)),
\end{equation}
where $d$ is the effective growth dimension.

\begin{figure}[h]
	\centering
	\includegraphics[width=0.4\textwidth]{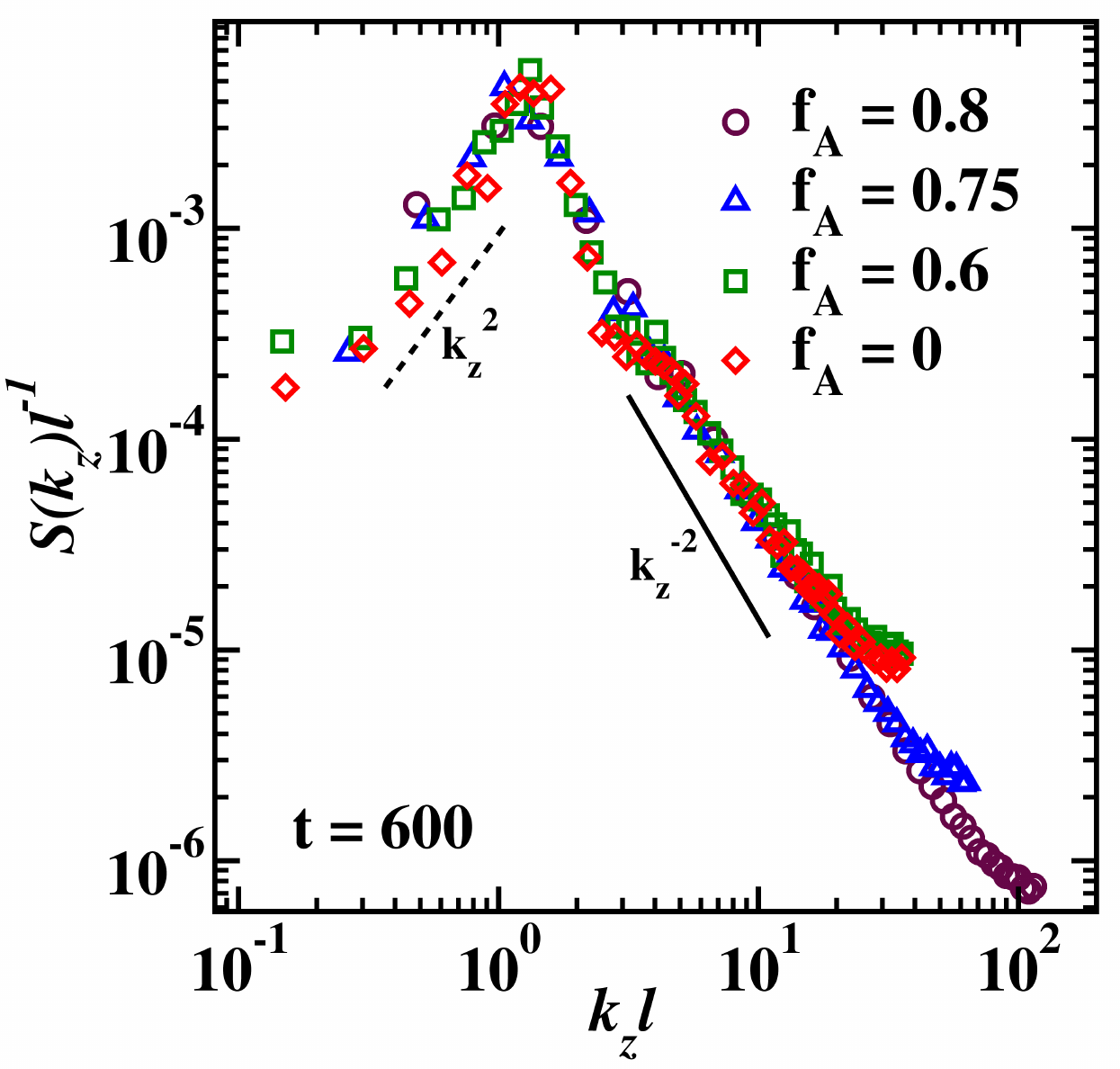}
	\caption{Scaling collapse of the structure factor $S(k_z,t)$ for passive and active systems. The large-$k_z$ tail follows a one-dimensional Porod law $S(k_z) \sim k_z^{-2}$ represented by solid line.}
	\label{fig:C_struct1}
\end{figure}

Figure~\ref{fig:C_struct1} shows satisfactory scaling collapse of $S(k_z,t)$ for both passive and active cases at a fixed time. At large $k_z\ell$, the structure factor exhibits a Porod-law decay, $S(k_z,t) \sim k_z^{-(d+1)} \sim k_z^{-2}$, consistent with one-dimensional ($d=1$) growth under cylindrical confinement~\cite{daniya}.  This observation further reinforces the validity of superuniversality.

\begin{figure}[h]
	\centering
	\includegraphics[width=0.4\textwidth]{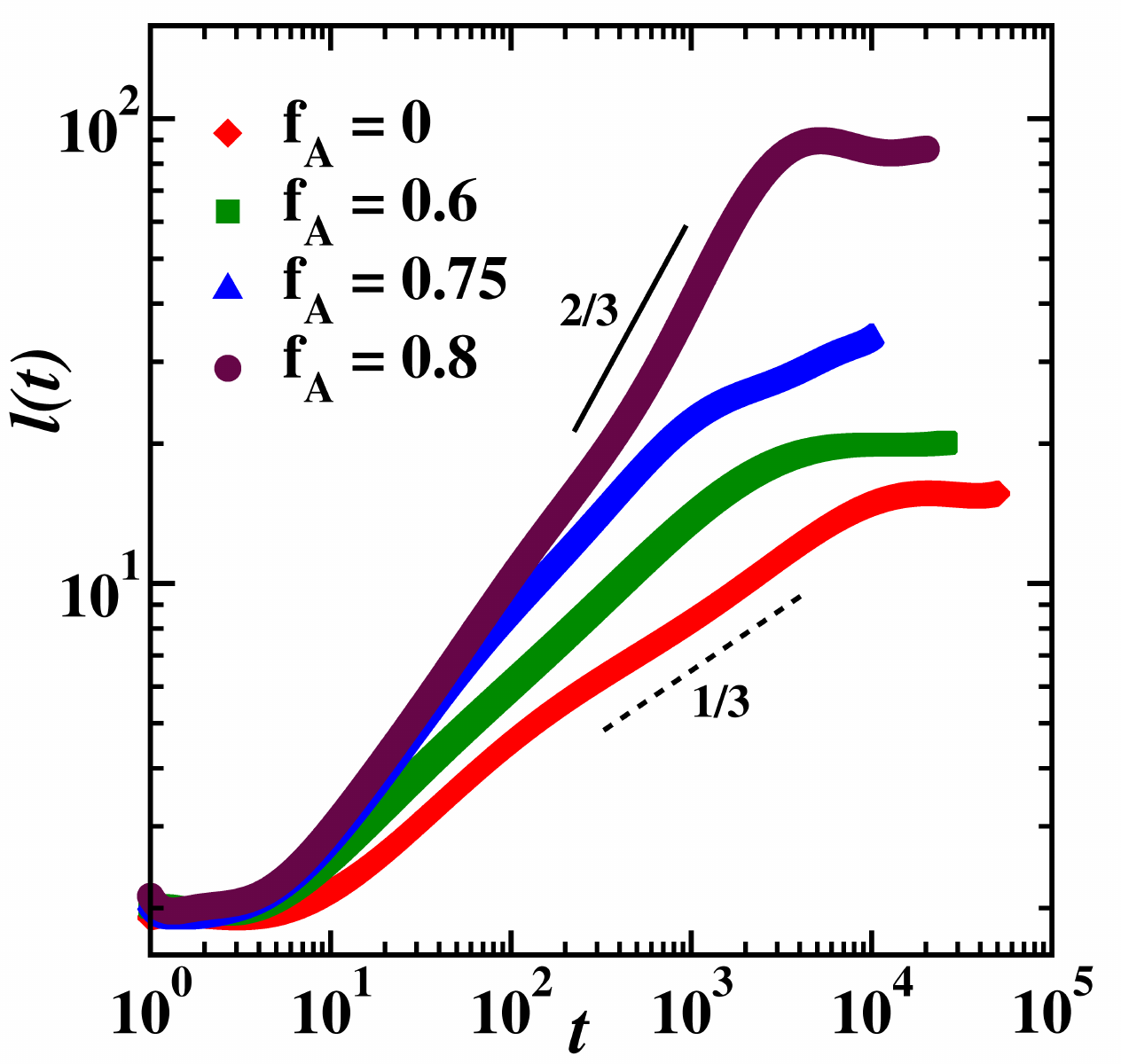}
	\caption{Growth of the characteristic domain size $\ell(t)$ for different activity strengths $f_A$. Activity induces a crossover from diffusive $\alpha = 1/3$ (dashed line) to ballistic $\alpha = 2/3$ (solid line) coarsening.}
	\label{fig:sfgsg}
\end{figure}

The coarsening kinetics is conveniently summarized by the time dependence of the characteristic domain size $\ell(t)$. Figure~\ref{fig:sfgsg} shows $\ell(t)$ as a function of time on a log-log scale for representative activity strengths. Both passive and active systems exhibit power-law growth,
\begin{equation}
	\ell(t) \sim t^{\alpha}.
\end{equation}
In the passive case ($f_A=0$), the use of a Langevin thermostat breaks local momentum
conservation due to stochastic particle-bath interactions. Consequently, coarsening proceeds via diffusive transport, yielding the classical Lifshitz-Slyozov growth exponent $\alpha=1/3$~\cite{lifshitz1961}. In contrast, increasing activity significantly enhances both the growth exponent and the saturation length scale. For $f_A=0.8$, after an initial diffusive regime, the system crosses over to a faster growth regime with $\alpha \simeq 2/3$. 

\begin{figure}[h]
	\centering
	\includegraphics[width=0.4\textwidth]{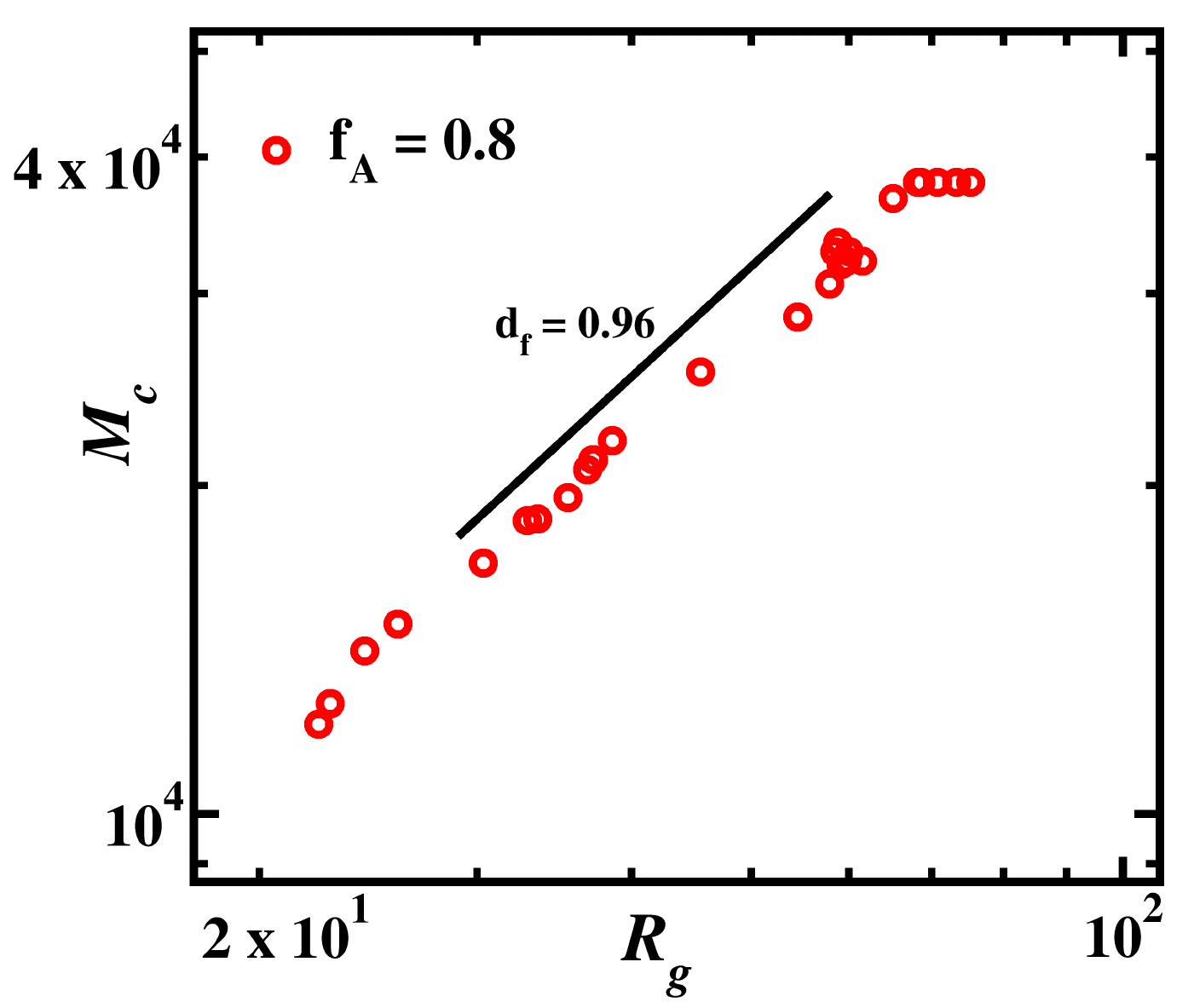}
	\caption{Average cluster mass $M_c$ versus radius of gyration $R_g$ for $f_A = 0.8$. A power-law fit (solid line) yields the fractal dimension $d_f \simeq 1$, consistent with quasi-1D coarsening.}
	\label{fig:c_m_vrms}
\end{figure}

This behavior can be understood within the framework of ballistic cluster coalescence~\cite{carnevale1990,trizac2003,paul2017ballistic}. As self-propulsion becomes effective, the dynamical behavior of the domains changes qualitatively compared to the passive case. In contrast to the arrested morphology observed without activity, the domains now acquire mobility. To describe this coarsening process, we consider a population of moving liquid clusters (or equivalently, domains) characterized by a time-dependent number density $n(t)$, which evolves due to cluster–cluster collisions and subsequent coalescence. We assume that this evolution can be described by the mean-field rate equation
\begin{equation}
	\frac{dn}{dt} = -\frac{n}{\tau},
\end{equation}
where $\tau$ denotes the mean collision time between clusters.
The cluster number density is related to the characteristic domain size $\ell(t)$ through the scaling relation $n \sim \ell^{-d}$, where $d$ is the effective dimensionality of the growing structure. The inverse collision time is estimated as $1/\tau \sim n v_{\mathrm{rms}} \ell^{d-1}$, where $v_{\mathrm{rms}}$ is the root-mean-squared velocity of the clusters and $\ell^{d-1}$ represents the effective collision cross-section. Combining these relations leads to
\begin{equation} \label{eq:17}
	\frac{dn}{dt} \sim -n^{(d+1)/d} v_{\mathrm{rms}},
\end{equation}
which explicitly shows how the coarsening rate is controlled by both cluster density and cluster mobility.
For ballistic cluster motion, the mean-squared displacement (MSD) grows quadratically with time,
\begin{equation}
	\langle r^2(t) \rangle = v_{\mathrm{rms}}^2 t^2 .
\end{equation}
To estimate $v_{\mathrm{rms}}$, we assume that the kinetic energy of a cluster of mass $M_c$ is governed by thermal fluctuations. Under this assumption, the equipartition theorem yields
\begin{equation}
	\frac{1}{2} M_c v_{\mathrm{rms}}^2 = \frac{3}{2} k_B T ,
\end{equation}
from which we obtain
\begin{equation}
	v_{\mathrm{rms}} = \sqrt{\frac{3 k_B T}{M_c}} \Rightarrow
	v_{\mathrm{rms}} \propto M_c^{-1/2}.
\end{equation}
This inverse square-root dependence is a fundamental consequence of thermal motion and reflects the balance between kinetic energy and cluster mass. Since the average cluster mass scales inversely with the number density, $M_c \sim n^{-1}$, it follows that
$v_{\mathrm{rms}} \sim n^{1/2}$, a scaling relation that is directly verified by our simulation data, as shown in Fig.~\ref{fig:c_m_vrms}.
Substituting this relation into Eq.~\eqref{eq:17} and solving the resulting rate equation leads to the growth law
\begin{equation} \label{eq:18}
	\ell(t) \sim t^{2/(d+2)},
\end{equation}
which corresponds to ballistic cluster coalescence and is distinct from diffusive Lifshitz–Slyozov growth.

\begin{figure}[h]
	\centering
	\includegraphics[width=0.4\textwidth]{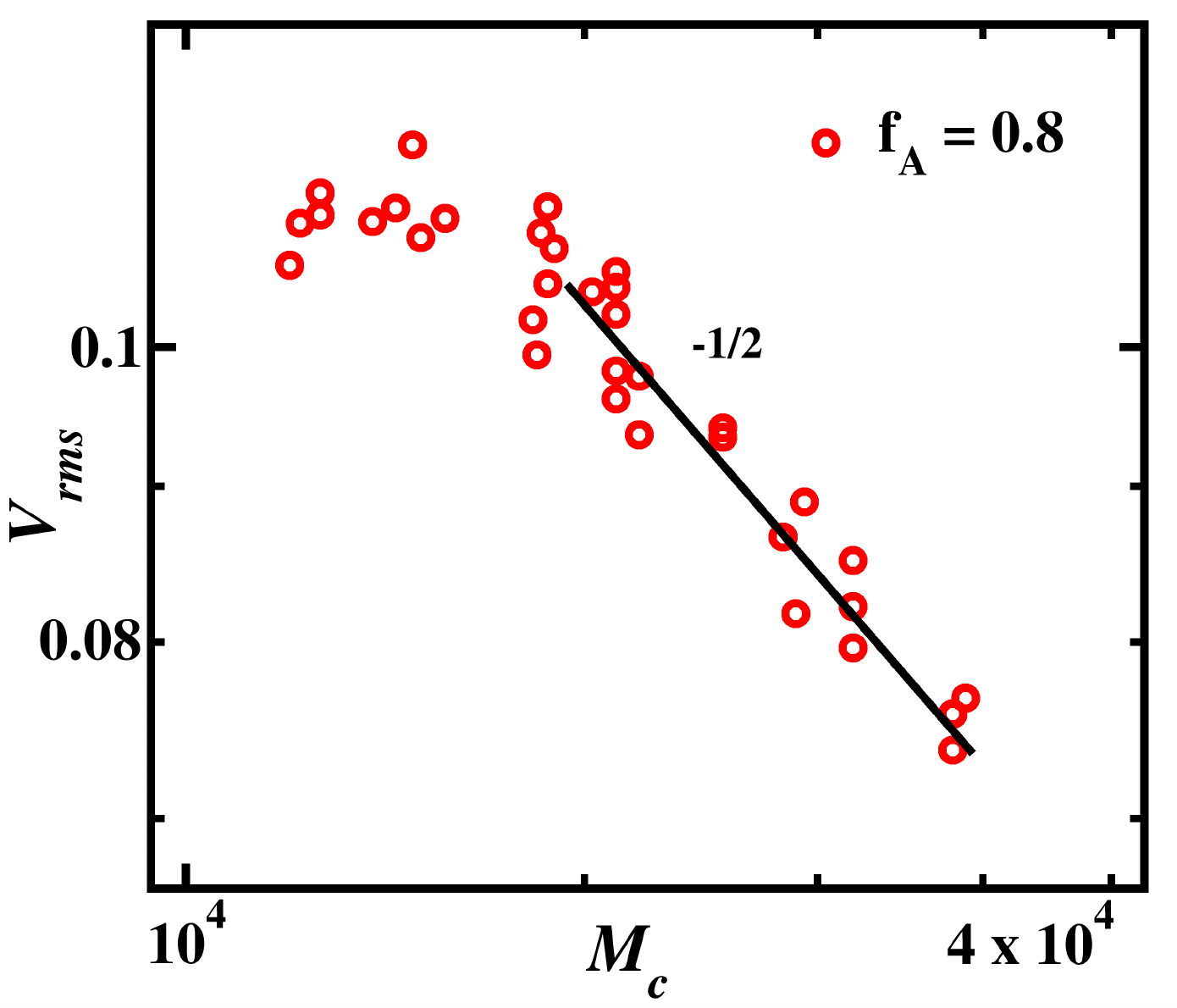}
	\caption{Root-mean-squared cluster velocity $v_{\rm rms}$ versus cluster mass $M_c$ for $f_A = 0.8$. The data follow $v_{\rm rms} \sim M_c^{-1/2}$ (solid line), consistent with ballistic cluster coalescence.}
	\label{fig:c_m_rg}
\end{figure}

To determine the appropriate effective dimensionality $d$ for the confined system, we compute the fractal dimension $d_f$ of the evolving liquid clusters using the scaling relation $M_c \propto R_g^{d_f}$~\cite{vicsek1992,matsushita1994}. Here, $R_g$ is the average radius of gyration, defined as
\begin{equation}
	R_g = \left\langle
	\left(\frac{1}{N_c}\sum_{i=1}^{N_c}
	(\vec{r}_i - \vec{R}_{\mathrm{cm}})^2
	\right)^{1/2}
	\right\rangle ,
\end{equation}
where $N_c$ denotes the number of particles in a cluster at a given time. The corresponding scaling behavior is shown in Fig.~\ref{fig:c_m_rg}. Replacing $d$ by the measured fractal dimension $d_f$ in Eq.~\eqref{eq:18}, we obtain
\begin{equation}
	\ell(t) \sim t^{2/(d_f+2)}.
\end{equation}
For the present quasi-one-dimensional confined geometry, we find $d_f \approx 1$, which naturally leads to the growth exponent $\alpha = 2/3$, in excellent agreement with the observed late-stage coarsening dynamics.
Finally, we note that the apparent saturation of $\ell(t)$ at late times for $f_A=0.8$ arises from finite-size effects. For a fixed pore diameter $D$, the final average domain width becomes independent of the pore length $L$ once $L$ is sufficiently large. This behavior is observed for both passive and active systems and reflects the intrinsic confinement-imposed limit on the maximum achievable domain size.

\section{Summary and Conclusions}
We have studied vapor-liquid phase separation of an active LJ fluid confined inside a cylindrical pore using molecular dynamics simulations with Vicsek-type alignment interactions. Our primary objective was to understand how activity modifies the morphology and coarsening kinetics of phase separation under strong quasi-one-dimensional confinement.

In the passive limit, following a quench below the critical temperature, the system initially undergoes spinodal decomposition but subsequently evolves into a periodically modulated, plug-like morphology along the pore axis. The resulting late-stage dynamics is consistent with diffusive growth characterized by the Lifshitz-Slyozov exponent $\alpha = 1/3$. At late times, coarsening becomes kinetically arrested, and the system remains trapped in a metastable striped state. This behavior arises from suppressed mass transport and the absence of effective bridging events between neighboring domains under confinement. 

The introduction of Vicsek-type activity qualitatively alters this scenario. Activity enhances particle transport and induces coherent motion of liquid domains along the pore axis, destabilizing the arrested striped morphology. For sufficiently strong activity, domain collisions and mergers become frequent, enabling complete phase separation even under strong confinement. Correspondingly, the late-stage coarsening exhibits a clear crossover from diffusive to faster, ballistic growth with an effective growth exponent $\alpha \simeq 2/3$, consistent with an advection-dominated cluster coalescence mechanism.

Analysis of the correlation function and structure factor shows that dynamic scaling holds across the activity strengths, while activity systematically modifies the domain morphology and transport properties. Despite these changes, the large-wave-number behavior remains consistent with one-dimensional Porod scaling, confirming
that confinement continues to control the effective dimensionality of coarsening.

Overall, our results demonstrate that alignment-induced activity can overcome confinement-induced kinetic arrest in vapor-liquid phase separation, leading to qualitatively new coarsening regimes. These findings provide insight into
nonequilibrium phase separation in confined active fluids and are relevant to a wide range of systems, including active matter in porous media and confined biological environments.

\textit{Acknowledgements.}--B. Sen Gupta acknowledges Science and Engineering Research Board (SERB), Department of Science and Technology (DST), Government of India (no. CRG/2022/009343) for financial support. Parameshwaran A. acknowledges DST-SERB, India for doctoral fellowship.

\end{document}